\documentclass[manuscript]{aastex}
\usepackage{graphicx}
\usepackage{txfonts}

\slugcomment{}

\begin{document}


\title{VISIR/VLT and VLA joint imaging analysis of the circumstellar nebula
 around IRAS~18576+0341}


\author{C. S. Buemi, G. Umana,  C. Trigilio and P. Leto}
\affil{INAF-Osservatorio Astrofisico di Catania, Via S. Sofia 75, 95123 Catania, Italy
}

\and

\author{J. L.  Hora}
\affil{Harvard-Smithsonian Center for Astrophysics, 60 Garden St. MS-65, Cambridge, MA 02138-1516}




\begin{abstract}
High spatial and sensitivity images of the Luminous Blue Variable IRAS~18576+0341
were obtained
using the mid infrared imager VISIR at the Very Large Telescope and the  Very Large Array interferometer. 
The resulting mid-infrared continuum maps show a similar clumpy and  
approximately circular symmetric nebula, which contrasts sharply with the 
asymmetry that characterizes the ionized component of the envelope, 
as evidenced from the radio and [\ion{Ne}{2}] line images obtained with comparable spatial resolution. 
In particular, there is excellent overall agreement between the 12.8 $\mu$m map and the radio images, 
consistent with free-free emission from circumstellar ionized material surrounding a central stellar wind.\par
The color temperature and optical depth maps obtained from mid-infrared images
show only slight fluctuations, suggesting quite uniform dust characteristics
over the dust shell. \par
We explore various possibilities to understand the cause of the different morphology 
of the dusty
and gaseous component of the circumstellar envelope which are compatible with the observations.

\end{abstract}




\section{Introduction}

During the  latest phases of their evolution, high mass stars experience significant mass loss, both through strong stellar winds and eruptive events, that drives these objects out of the main sequence through short and poorly constrained instability phases. 
The class of Luminous Blue Variables (LBVs) consists of luminous ($\log$(L/L$_{\odot}\,\geq\,5.5 $) and massive
 stars that are believed to go through a short 
but violent transition from the main sequence towards the Wolf-Rayet stage \citep{HD94}, although their link to other advanced evolutionary phases of massive stars such as supernovae is still open \citep{Barlow:2005, Kotak:2006}.
During their evolution, LBVs lose a huge quantity of mass from their original envelope, leading to the formation of extended circumstellar nebulae (LBVN), whose masses are thought to be on the order of a few solar masses.
In the galactic context, despite the relative small number of confirmed and candidate members, LBVs represent a  significant stage of the cosmic cycle of matter, as they contribute to the chemical enrichment of interstellar medium in dust and heavy elements \citep{Smith:2006}. The census of Galactic LBVs performed by \cite{clark05} reported 12 effective members and 23 candidates, but new candidates have recently been identified via the mid-infrared images of the Galactic plane obtained with the Multiband Imaging Photometer for Spitzer \citep[MIPS;][]{rieke04} and the Infrared Array Camera \citep[IRAC;][]{fazio04} onboard the Spitzer Space Telescope \citep{wachter10, gvaramadze10a, gvaramadze10b}.

The overall scenario of LBV evolution is currently under debate and the nebula formation mechanism, i. e. intensity, duration and geometry of the mass-loss events, is not well established.
The study of the gas and dust content of the  nebula  which surrounds the star may provide significant clues that will enable us to understand the mass-loss processes that govern its evolution, as well as the drastic changes that take place in the immediate circumstellar environment \citep{Umana09, Jimenez10}.
By exploring the characteristic of the mid infrared continuum emission, which traces the warm dust,
and of the recombination lines and radio continuum emission, which trace  the ionized gas, 
we can study the relationship between these different components which coexist in the stellar ejecta with the aim of gaining valuable insights about the formation and shaping mechanisms at work in the circumstellar envelope.  In particular, radio observations offer the additional possibility  
to reveal the ionized gas well inside the dusty envelope and, therefore, allow to determine in great detail its spatial distribution without suffering from intrinsic extinction \citep{Umana05, Umana10}.\par
 
The infrared source \object{IRAS~18576+0341} was originally classified as a Planetary Nebula on the basis of its infrared colors \citep{garcialario} and of the observed 1.4 GHz  flux \citep{Condon99}. Subsequently, the detection of mid-infrared extended emission around this object \citep{ueta01}, together with observations of strong photometric and spectroscopic infrared variability
\citep{ueta01, Pasquali_02, clark_03, clark09} and its position in the HR diagram led to a definitive classification of  \object{IRAS~18576+0341} as a LBV.\par 
On the basis of radiative transfer calculations, \cite{ueta01} 
derived a $\log {\frac{L}{L_{\odot}}}$~=~6.4 and a temperature $T_{eff} = 15\pm 9$ kK for the central star, which appears surrounded by a nebula of gas and dust characterized by a possible C/O mixed chemistry. 
Recently, \cite{clark09}, in a detailed analysis of long term monitoring of the near infrared photometric and spectroscopic 
properties of  \object{IRAS~18576+0341}, showed that 
it has been highly variable over the last 20 years. Near-infrared photometric variability greater than  one magnitude has been observed with moderate variation in temperature ($\Delta T_{*}\leq 4.5 $kK),
which is consistent with variations of stellar radius.  \par 

Multi-frequency VLA observations have revealed an extended, asymmetric, and quite structured nebula and revealed the central core of the LBV, whose radio properties are typical of a stellar wind, with a  current mass-loss of $3.7 \times 10^{-5} M_{\odot}\ yr^{-1}$ \citep{Umana05}. This value is consistent with those determined by \cite{clark09} from the NLTE atmospheric code, which indicates a variation of the mass loss rate
from $1.2 \times 10^{-4}$ to  $5.2 \times 10^{-5} M_{\odot}\ yr^{-1}$ between 2002 and 2006, associated with a decrement of $T_{eff}$ from 15.5 to 11.0 kK.\par

The present work is a part of an effort to better understand the physical conditions in a sample of Galactic  LBV nebulae, conducted by using state of the art instruments available at mid infrared and radio wavelengths. 
We present the results obtained for the LBV IRAS~18576+0341 by combining mid infrared
and radio maps having comparable spatial resolution, obtained using VISIR at Very Large Telescope (VLT) and Very Large Array (VLA). The images allow us to study the detailed morphology of the circumstellar
envelope, in order to understand the origin of the apparently contradictory spatial structures seen in the 
resulting maps \citep{Umana10}.

\section{Observations and Data Reduction}
\subsection{VISIR/VLT Observations}
To map the dust distribution in the LBVN associated with IRAS~18576+0341, we have obtained  high angular resolution and high sensitivity N and Q-band  images with VISIR  \citep{Lagage2004}, the VLT imager for the mid-IR mounted at the Cassegrain focus of the VLT Unit 3 telescope (MELIPAN). The observations were carried out on 2007 July 19 and 26 through the PAH2 filters,
 centered on known PAH features ($\lambda_{c}=11.26 \mu m, \Delta\lambda=0.59 \mu m$) and adjacent continuum (PAH2$_2$,
$\lambda_{c}=11.88 \mu m$, $\Delta\lambda=0.37 \mu m$), and in Q1 ($\lambda_{c}=17.65 \mu m$, $\Delta\lambda=0.83 \mu m$).
We also used the NeII filter ($\lambda_{c}=12.80 \mu m$, $\Delta\lambda=0.21 \mu m$) and its adjacent continuum (NeII$_2$ $\lambda_{c}=13.03 ~\mu m$, $\Delta\lambda=0.22 ~\mu m$) as a tracer of ionized gas to compare with the radio images. All the observations were performed under very good and stable weather, with an optical seeing of about $0.7\arcsec$. The target was observed at airmasses ranging between 1.2 and 1.4.
We  used a fixed pixel scale of $0.075\arcsec$, resulting in a FOV of $19.2\times 19.2 \arcsec$. The standard chopping/nodding technique was adopted for subtraction of the sky background as well as the telescope's thermal emission: secondary mirror chopping was performed in the North-South direction, with a chop throw of 
$9\arcsec$ and telescope nodding was applied in the opposite direction with equal amplitude.
To further improve the image quality, a random jitter pattern with a maximum throw of $3 \arcsec$ was superimposed on the nodding sequence.  A complete log of observations is summarized in Table~1.
Data were reduced following the standard VISIR pipeline (version 1.7.0), consisting of co-adding of frames after flat fielding and removal of bad pixels. The chopped and nodded images were then combined to make one image in each filter. 
Three mid-IR standards were observed just before and after the target acquisition to flux calibrate the data.
Such observations were also used to determine the FWHM of standard star images and thus to derive the actual angular resolution of our final maps. Values of the FWHM for each filter are reported in Table~1.
\subsection{VLA observations} 
We observed IRAS~18576+0341 on the 2004 October 12  at 4.8 GHz (6cm) and 1.4 GHz (20 cm), with a bandwidth of 100 MHz using the VLA in A configuration,
providing a typical beam size of $\approx$0.36$^{\prime\prime}$ and 
$\approx$1.3$^{\prime\prime}$ respectively. For all the observed bands, 1824+107 was used as
phase calibrator and the flux density scale was determined by observing 3C~48.

The data  processing was performed using the standard programs
of the NRAO {\bf A}stronomical {\bf I}mage  {\bf P}rocessing  {\bf S}ystem (AIPS).
We compared the images made from the A-configuration data with our earlier images obtained using the C configuration \citep{Umana05} to confirm that long term variability was not present in our data.

After the calibration process, the A-configuration {\it uv}-datasets were combined
with the earlier observations performed at the same frequencies 
using the VLA in C configuration and presented in \cite{Umana05}
and images have been produced at both wavelengths.
This allowed us to obtain high resolution maps of the entire nebula and, in the same time, fully recover all the extended  
emission.
The mapping process was performed by using AIPS task IMAGR with a variety of 
weighting schemes and the dirty maps were {\bf CLEAN}ed down as close as possible to the theoretical noise. The final uniform-weighted (ROBUST -5) maps have a synthesized beam of 0\farcs38$\times$0\farcs36 and 1\farcs39$\times$1\farcs27,
 and rms noise of 0.04 mJy~beam$^{-1}$ and 0.1 mJy~beam$^{-1}$ at 6~cm and 20~cm, respectively.
The noise level (rms) in the maps  was estimated by analyzing an area on the map
(using task IMEAN),
whose dimension is of the order of more than 100 $\theta_\mathrm{syn}^{2}$
away from the phase center and free from evident radio sources.

\section{Results}
\subsection{The dust shell}
\begin{figure}
\resizebox{14cm}{!}{\includegraphics{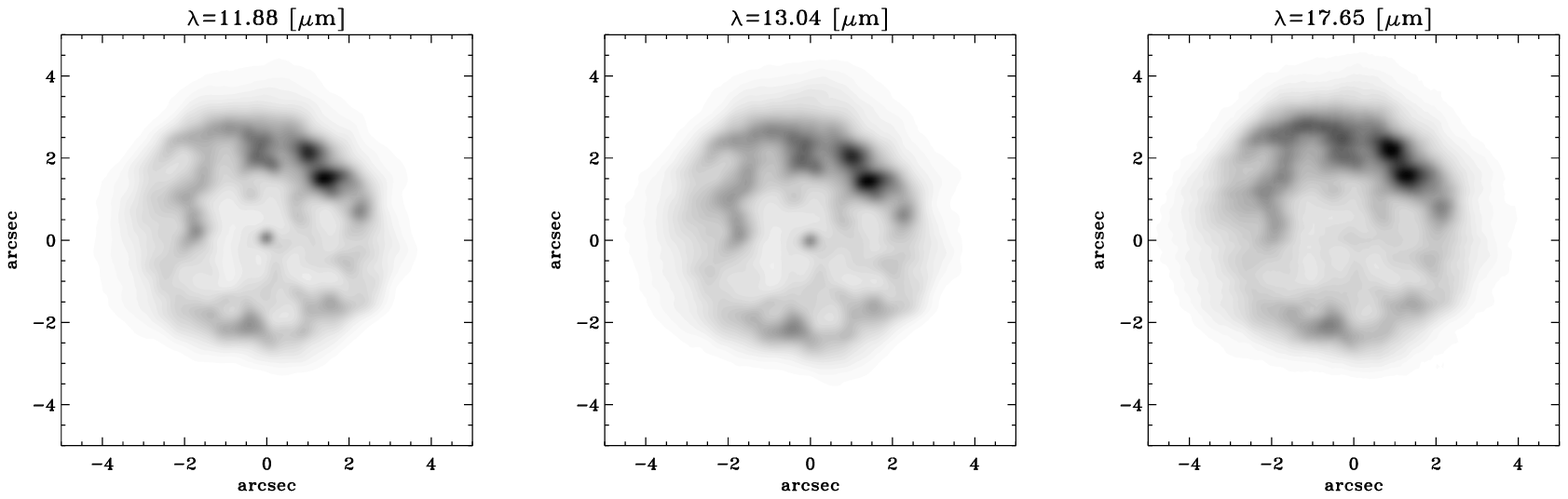}}
\caption{VISIR images of IRAS~18576+0341 and its surrounding nebula taken through the filters
PAH2$_2$, NeII$_2$ and Q1.}
\label{ir_im}
\end{figure}

All five VISIR images, shown in Figure~\ref{ir_im}, have a similar morphology, with an extended circumstellar envelope surrounding a central source that is clearly visible until almost 13 $\mu$m. In Figure~\ref{ir_im} we show the images
obtained in the filters PAH2$_2$, NeII$_2$ and Q1. 
From all the filters we derived a shell diameter of approximately 7$^{\prime\prime}$, corresponding to a size of 0.35~pc  at a distance of 10~kpc \citep{clark09}.

We performed aperture photometry of the central object in each image by using the IDL procedure ATV,
while the nebula contribution at each wavelength has been determined by integrating the
flux emitted in a selected area of $ 9 \arcsec \times 9 \arcsec$. 
The photometric results are summarized in Table~\ref{visir-phot}.
On the basis of their ISO SWS spectrum, \cite{Hrivnak00} pointed out a strong difference between the IRAS and the simulated ISO photometry, concluding that the dusty envelope around IRAS~18576+0341 should have been very extended with some of the flux observed by IRAS falling outside of the largest ISO aperture ($ \geq 30^{\prime\prime}$).
However, we fully recover the IRAS 12 $\mu$m flux density (F$_{60 \mu m}$=58.48 mJy), implying that, at least at 12 $\mu$m, all of the emission is from the compact 7$^{\prime\prime}$ source. 

With respect to the previous mid-infrared images \citep{ueta01} that discovered the presence of a roughly circularly symmetric nebula around IRAS~18576+0341, our higher resolution and sensitivity allow us to probe finer details of the dust distribution. 
The brightness of the shell varies around its circumference, with the major contribution to the flux coming from the north-west part
of the nebula. The condensations, detected by \cite{ueta01} as an arc located in the north side of the shell, now shows up as a clumpy structure with two major peaks located at 50 degrees in the East-North direction and about 
2$^{\prime\prime}$ from the central star.
It appears that the northern and the southern emission peaks reported by \cite{ueta01} were an effect of the limited 
sensitivity and spatial resolution of their maps.
From our VISIR maps, no hints for the claimed optically thin edge-on dust torus surrounding the central object are evident, but only a suggestion of a possible spherical  inner shell whose edges are strongly structured.  

\begin{figure}
\resizebox{10cm}{!}{\includegraphics{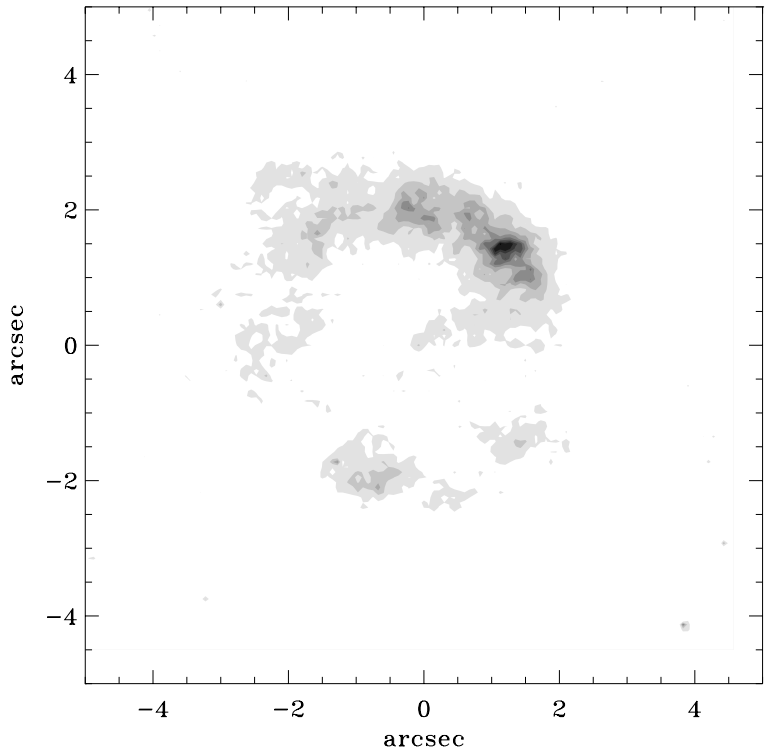}}
\caption{Continuum subtracted 11.26  $\mu$m image of IRAS~18576+0341,
showing the PAH feature emission.}
\label{PAH}
\end{figure}

It is possible to isolate the PAH feature at $11.26\, \rm{\mu m}$ by subtracting the image
obtained at the adjacent continuum (PAH2$_2$) from the the PAH2 image. We have performed this subtraction after registering
the images on the central object and normalizing them to the continuum as derived from a fit to the ISO spectrum  \citep{Sloan03}.  The difference image is shown in Figure~\ref{PAH}.
The $11.26\, \rm{\mu m}$ PAH emission appears to be located in the inner shell with an higher
concentration in the north-west part of the ring where the most of the thermal dust continuum emission originates.

\begin{figure}
\resizebox{10cm}{!}{\includegraphics{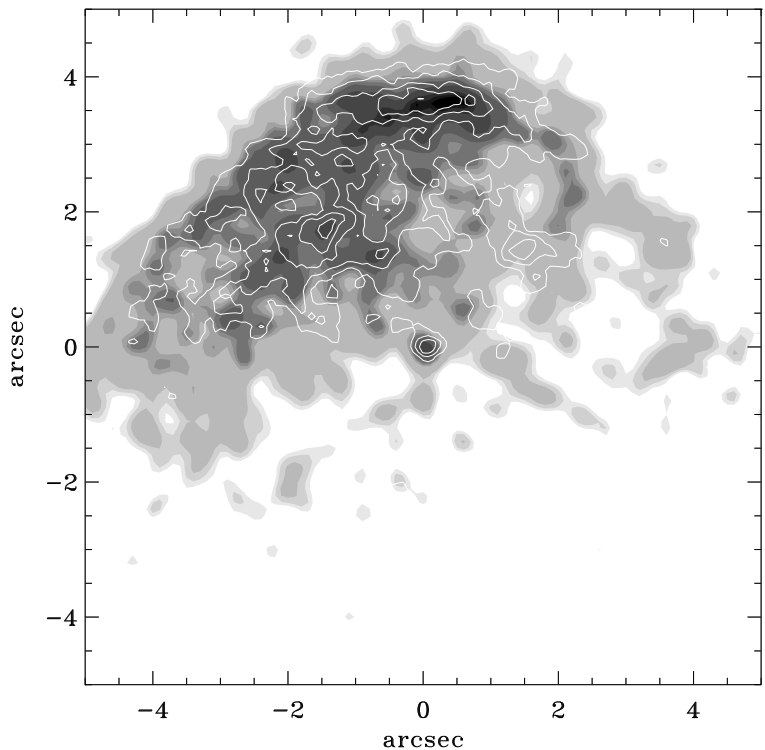}}
\caption{Continuum subtracted 12.8 $\mu$m contour map of IRAS~18576+0341,
showing the [Ne~II] line emission, superimposed to the 6 cm image.}
\label{NeII}
\end{figure}

The [\ion{Ne}{2}] emission, which should trace the ionized gas, can be isolated
following the same procedure, i. e., by subtracting the adjacent continuum image at
$13.03\,\rm{\mu m}$ (NeII$_2$) from the $12.80\,\rm{\mu m}$ image. 
The striking result, shown in Figure~\ref{NeII}, demonstrates that the [\ion{Ne}{2}] emission has a
very different distribution compared to the dust and PAH tracers.
As [\ion{Ne}{2}] emission is a ionized gas tracer this result suggests that significant difference exist between the dusty 
and ionized stellar ejecta (sec.~\ref{ion_gas}).

Assuming that the extinction is constant across the images and that the flux that we have mapped is due to optically thin thermal emission from dust grains,
we may use the NeII$\_$2 and Q1 images to produce a ratio map.
In this case the observed intensity is given by
\begin{equation}
\label{trasporto}
I_{\nu}\approx B_{\nu}(T)\tau_{\nu}.
\end{equation}
and using the Wien approximation ($h\nu \gg kT$) for the blackbody expression $B_{\nu}(T)$,
the ratio of the intensities at two different frequencies $\nu_{1}$ and  $\nu_{2}$ is given by:
\begin{equation}
\frac{ I_{\nu_1} }{I_{\nu_2} }=e^{-h(\nu_1-\nu_2)/kT}
\left(\frac{ \lambda_{2}}{\lambda_{1}}\right)^{3}\frac{\tau_{\lambda_1} }{\tau_{\lambda_2}}
\label{rapporto}
\end{equation}
The optical depth $\tau_{\lambda}$ depends on the chemical composition and on the size $a$ of the grains.
There is an open controversy on the nature of dust grains present in the nebula 
surrounding IRAS~18576+0341 \citep{ueta01, Hrivnak00}, indicating that it is possibly a mixed-chemistry object where features due to silicates and those related to carbon (PAH) coexist.
In this paper, we follow \cite{Umana10} and adopt 
for $\lambda_1=11.88$ and
$\lambda_2=17.65\,\rm{\mu m}$, $\tau_{\lambda_1}/\tau_{\lambda_2}=1.05$ in the
case of graphite and 1.27 in the case of astronomical silicates.

For each pixel of the map, we derive the dust temperature inverting equation~\ref{rapporto} 
\begin{equation}
T\approx 1.44\times 10^4 
\frac{(\lambda_2^{-1} - \lambda_1^{-1})}
{\ln\left[ \frac{ I_{\nu_1}} {I_{\nu_2}} 
\frac{\tau_{\lambda_2} }{\tau_{\lambda_1}}
\left(\frac{\lambda_1}{\lambda_2}\right)^{3}
\right] }
\end{equation}

The dust temperature has been computed only at those pixels whose brightness is greater
than $0.06\,\rm{Jy\, arcsec^{-2}}$, corresponding to $3\,\sigma$ in the maps. In this
derivation, we assume a constant temperature along the line of sight for each pixel.

\begin{figure}
\resizebox{17cm}{!}{\includegraphics{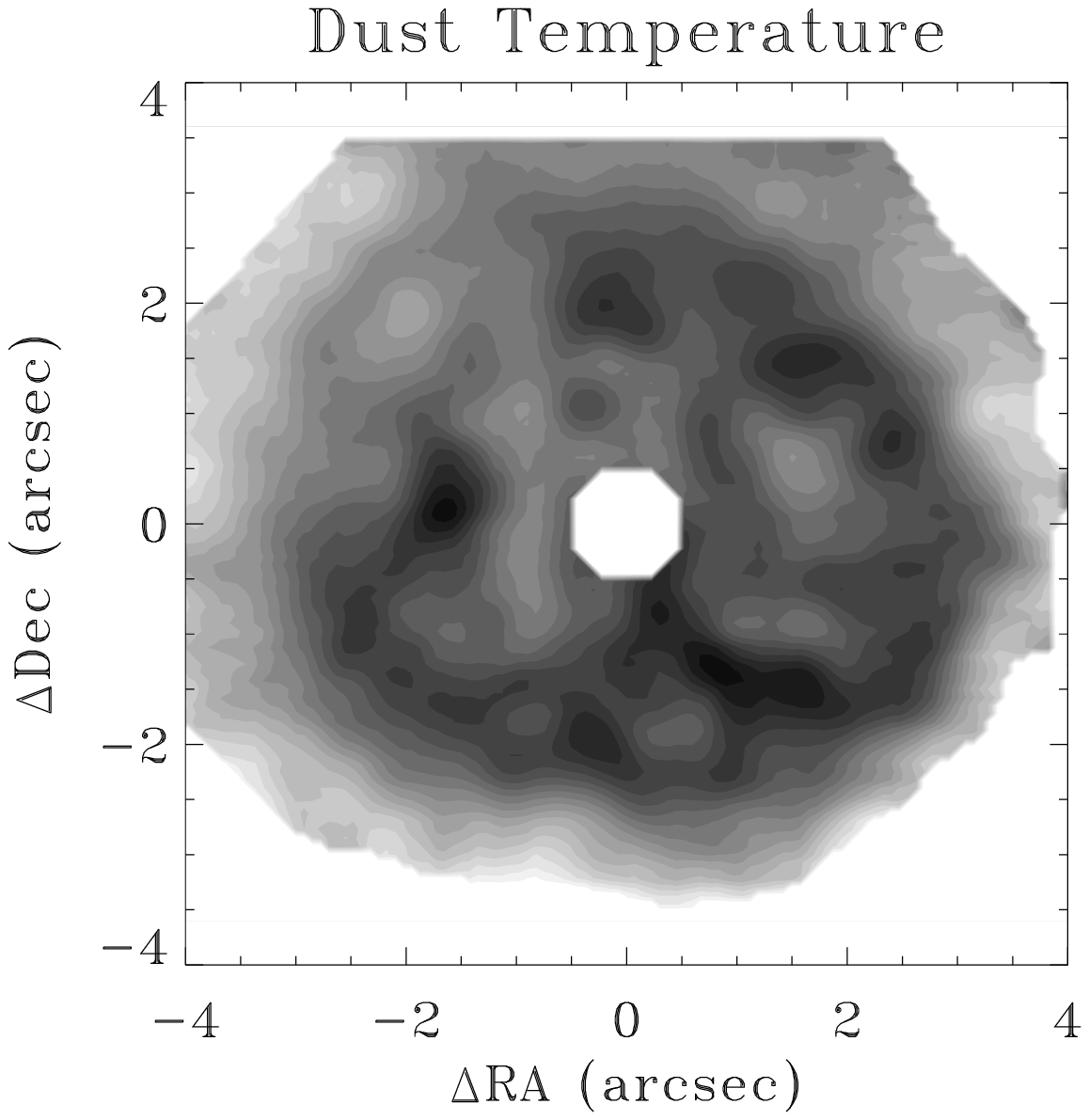}\includegraphics{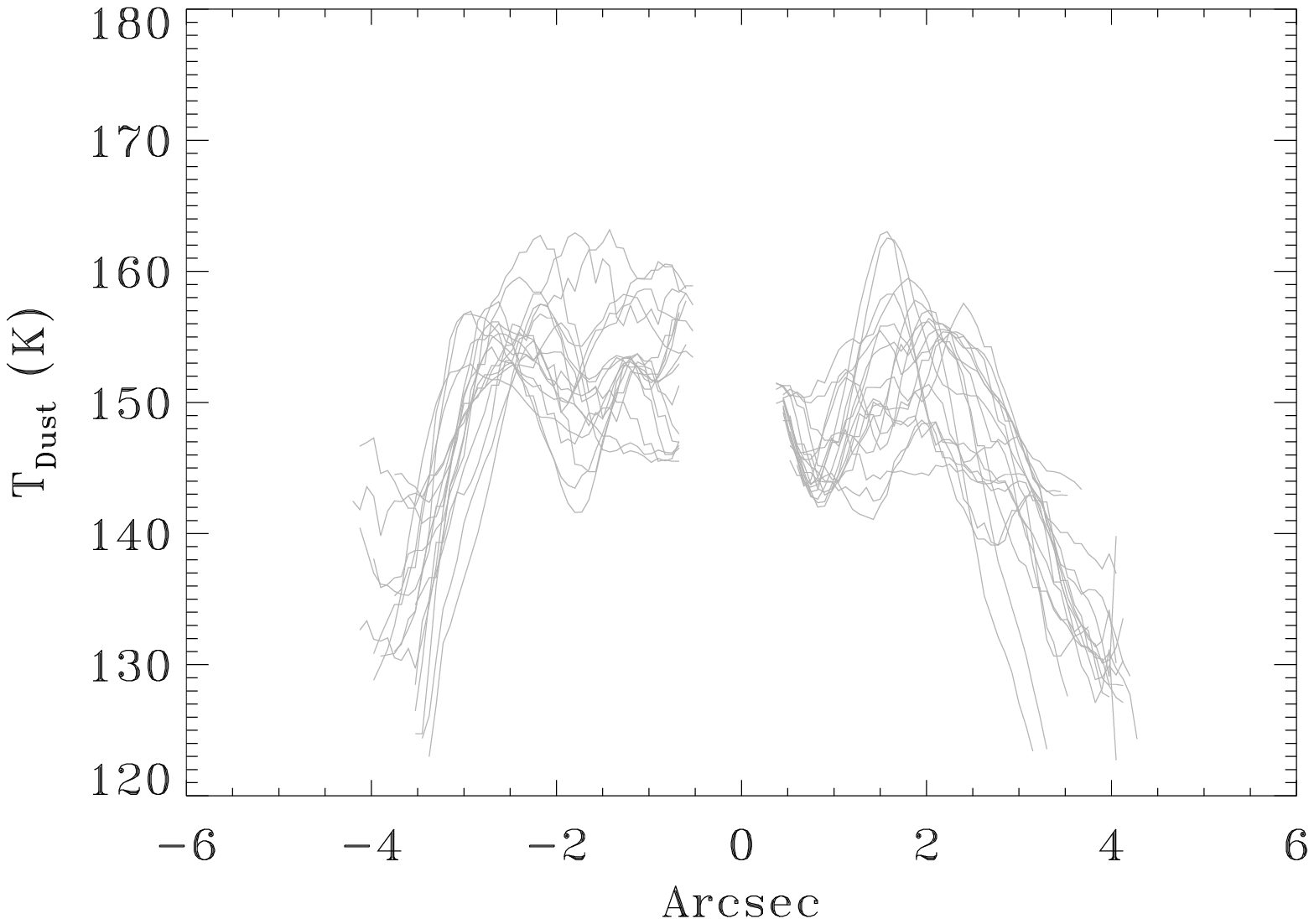}}
\caption{Left panel: map of dust color temperature as determined from 
$\lambda_1=11.88$ and
$\lambda_2=17.65\,\rm{\mu m}$ maps. For the dust a Carbon chemistry has been assumed. Regions
of warmer and cooler emission are seen as darker and lighter shades respectively.
Right: Temperature profiles along diameters at step of 10\degr; here
the central part with the star is not considered; the higher values are reached at about 2\arcsec~
from the center.
}
\label{tempmap}
\end{figure}

The resulting dust temperature map is shown in the left panel of Figure~\ref{tempmap}.
The temperature map can be used to examine dust temperature gradients across the nebula. 
We see that the color temperature values are confined to a relatively narrow range, 
from $\sim$ 130 to 160 K in case of graphite grains.
Slightly lower values ($\sim 10\%$)
are obtained if instead a mixture of silicate grains is assumed.

The higher temperatures are reached at a distance of about 2\arcsec~ from the center. 
To more clearly see the temperature distribution as a function of radius, in the right panel of Figure~\ref{tempmap} we show radial cuts of the temperature map along diameters obtained at steps of $10\degr$.

To evaluate the mass of the dust in the envelope,  an optical depth map at $\lambda=17.65\,\rm{\mu m}$  is required.
This can be obtained by inverting Eq.\ref{trasporto}, using the value of the temperature found in each point of the map and the observed intensity.  
If the medium is homogeneous, the optical depth is given by 
$\tau=k_{\rm dust}\, \rho_{\rm dust}\, l$, where $\rho_{\rm dust} $ 
and $l$ are the the density of the medium and the thickness of the envelope, and $k_{\rm dust}$ is the absorption coefficient per mass unit. 

The mass behind each optical depth map pixel, whose area is 0\farcs075 $\times$ 0\farcs075 , is given by
$\Delta m=\tau/k\times \Delta A$, and $\Delta A$ the area source corresponding to one map pixel, which is $1.25 \times 10^{32}\,\rm{cm^{2}}$ assuming a distance of 10~kpc \citep{clark09}.
A value of $k_{\rm dust}=253$ and $1082\,\rm{cm^2\,g^{-1}}$ at $\lambda=17 \,\rm{\mu m}$  has been adopted 
in case of graphite and silicates, respectively  \citep{Umana10}.
Integrating over all the pixels with brightness higher than 3 $\sigma$, a total dust mass 
of $8.6\times 10^{31}\rm{g}$ ($ 4.3 \times  10^{-3}M_\sun$) is derived  in case of graphite, and 
$1.3\times 10^{32}\rm{g}$ ($6.5 \times 10^{-3}M_\sun$) in case of silicates.
\cite{ueta01} provide an estimate of total dust mass of $\sim 0.1M_\sun$ assuming
a toroidal morphology for the entire dust emitting region and a radius of $\sim$ 14$^{''}$. This value is compatible with our results
if scaled down to the extent of the mid-infrared emitting dust shell that is resolved in our images. 
Because the  mid-IR observations we present only trace the cooler dust component 
in the nebula, our dust mass estimates should be considered as lower limits to the total dust mass in the LBVN. 

\begin{figure}
\resizebox{17cm}{!}{\includegraphics{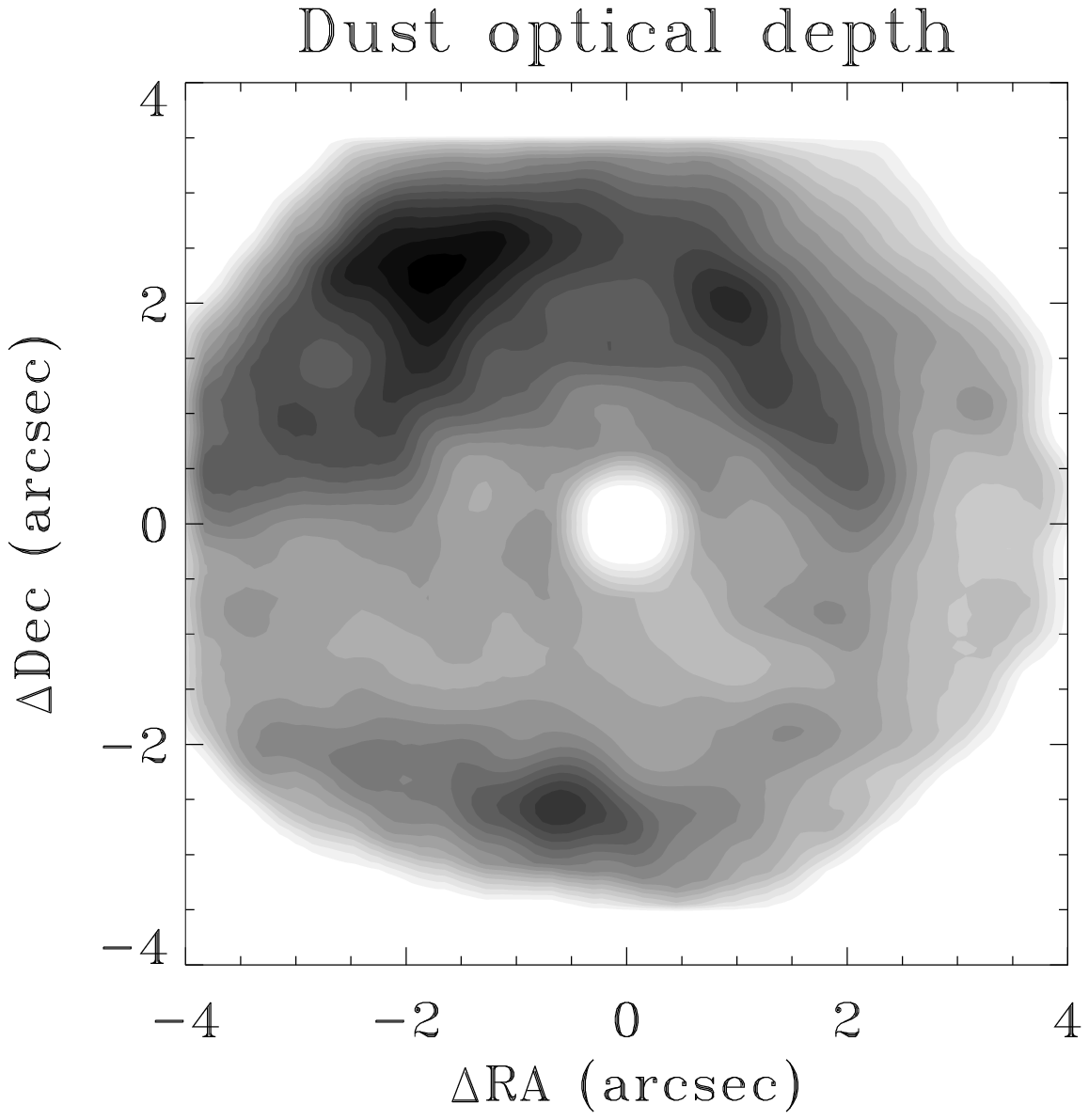}\includegraphics{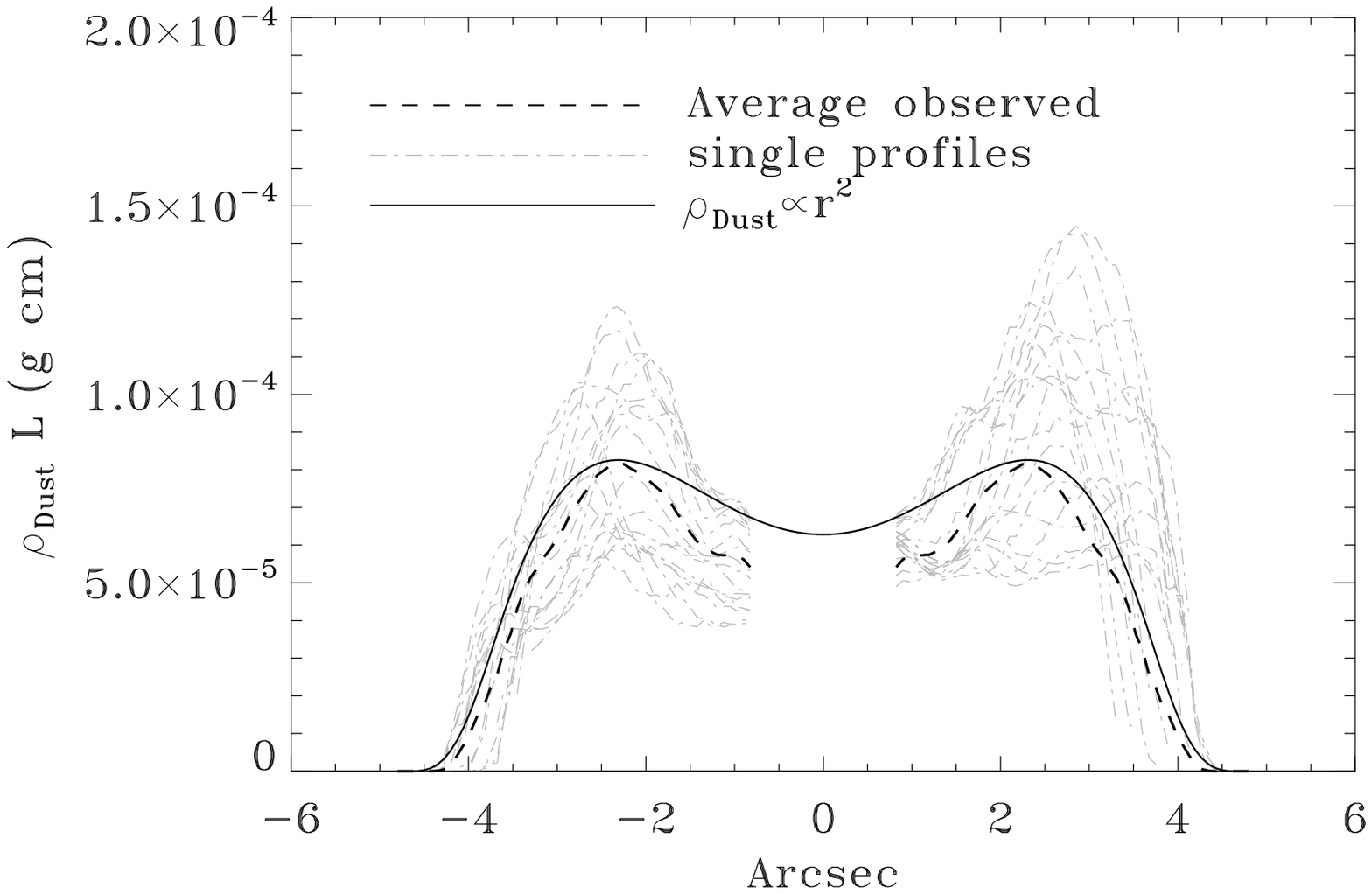}}
\caption{
Left: Map of dust the optical depth. The greatest concentration of mass is located in the north-east
part of the nebula.
Right: Column density profile: the average of the observed diagonal cuts is shown as dashed line; here
the central part with the star is not considered; a profile obtained with the density increasing
outward is shown as a continuous line.}
\label{taumap}
\end{figure}

The dust column density along the line of sight $\rho_{\rm dust}\, l$ (Figure~\ref{taumap}, left panel) 
can be used to infer the density $\rho_{\rm dust}$ inside the nebula. 
Although its distribution is clumpy, as already noted, 
with a higher concentration of dust in the north-east part, a roughly spherical symmetry around the central star can be assumed. Cuts of the column density map along diameters, obtained at steps of $10\degr$
averaged together, are shown as a thin line in the right panel of Figure~\ref{taumap}.
We modeled the density distribution assuming a radial dependence. For this purpose, we 
built a 3D matrix, sampled with a step corresponding to the pixel size of the VISIR maps 
($1.12\times 10^{16}$ cm) in all the three spatial dimensions. 
We found that the density must increase as a function of radius from the center, 
although it is not possible to discriminate among several radial dependencies.
As an example, the radial column density profile obtained assuming $\rho_{\rm dust}\propto r^2$,
with a smooth at the edge, is shown as a thick line in Figure~\ref{taumap}, right panel.

This analysis shows that in the inner part of the nebula the density is low and it increases
strongly with the distance from the star. This could indicate:\\
a) an episode of strong mass loss at a previous time;\\
b) the dust is processed by the UV radiation field of the star in the inner part of the nebula.\\
c) a combination of the two.\\
The maximum of $\rho_{\rm dust}$ is at $3\arcsec$ from the center, corresponding to a linear distance of
0.15~pc; assuming the same expansion velocity $v_{\rm{e}}\approx 70 \;\rm{km\,s^{-1}}$ 
measured from the Br$\gamma$ line \citep{clark09},
the mass loss episode should have occurred about 2\,000 years ago.

\subsection{Ionized gas}
\label{ion_gas}
\begin{figure}
\resizebox{14cm}{!}{\includegraphics{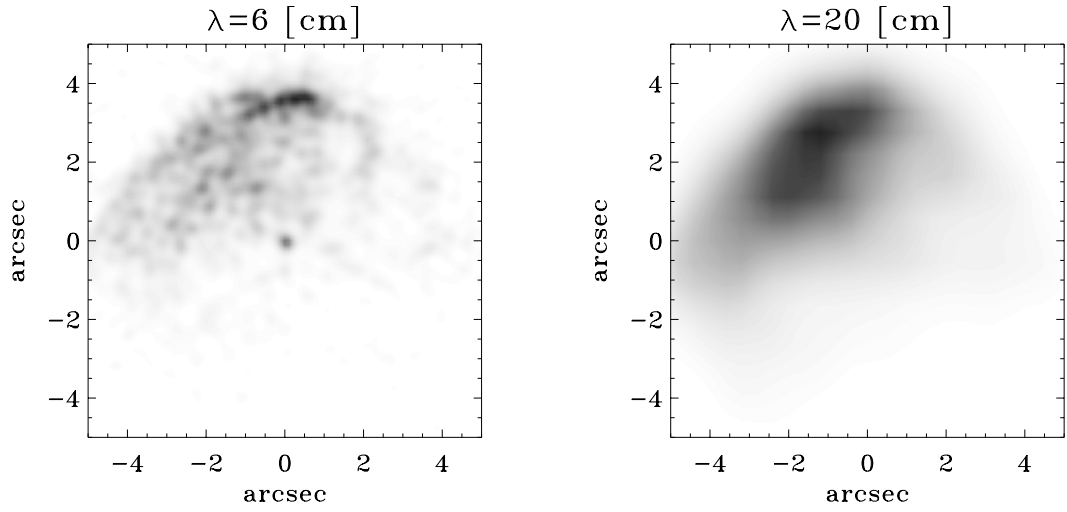}}
\caption{Uniform weighted radio maps of IRAS~18576+0341 at 6 and 20~cm, obtained by
merging A and C VLA configuration data.}
\label{radio_maps}
\end{figure}

The continuum emission associated with \object{IRAS~18576+0341} at 5 GHz and 1.4 GHz is shown in Figure~\ref{radio_maps}. 
The spatial resolution achieved in the final image is  about 1\farcs3 in the L Band and 0\farcs4 in the C Band.
The images confirm the structure of the source as detected at higher frequencies \citep{Umana05}: a compact, slightly
 resolved core, most likely associated with the stellar wind from the central object, surrounded by an extended ionized nebula with a remarkably similar shape and extension at all wavelengths. 
As expected, the combined effect of reduced spatial resolution and of intrinsic contribution at low frequency prevent us from detecting the core component and discerning it from the extended nebula in the 20 cm map;
on the contrary,  the higher spatial resolution in the 6~cm map allows us to clearly resolve the compact source. \par

 The total radio flux densities associated with the entire source have been derived
by summing the contribution of all the pixels contained in the radio source (AIPS task IMSTAT), while
the flux density of the central object object has been obtained by fitting a two dimensional Gaussian brightness
distribution to the map (AIPS task JMFIT). The results are reported in Table~\ref{tab_radio}.

To better separate the core component from the remaining diffuse emission, the flux density from the compact
core at 6 cm has been determined by fitting a two dimensional Gaussian brightness distribution to the map obtained with the A configuration data alone. The resulting flux density of 1.3$\pm$0.1 is in accordance with the flux density extrapolated from
measurements at higher frequencies assuming a  stellar wind spectrum, as reported in \cite{Umana05}.\par

The final images in Figure \ref{radio_maps} show that the radio emission from the circumstellar ionized nebula is 
asymmetric with respect the central core, extending out in the northwest direction, with the brightest radio region 
located about 3\farcs6 to the north of the central core in the 6 cm map.  The diffuse emission is remarkably
similar to that detected at all the other observing frequencies, with similar extent of about 7$^{\prime\prime}$.  
The southern component, which contributes about 25\% of the total mid-infrared emission, is not present in any of the radio maps. A certain degree of asymmetry is also present in the radio emission of other LBVNs, such as \object{HR~Car} \citep{white00} and \object{Pistol Star} \citep{Lang05, Lang99}. This may indicate that extremely asymmetric mass loss can be quite common in LBVs.  

The overall morphology of the radio nebula agrees very closely with the spatial distribution of the 
[\ion{Ne}{2}] line emission. The contours of the [\ion{Ne}{2}] map, superimposed on a gray-scale image of the 
6 cm data, is shown in Figure~\ref{NeII}. It is evident how the emission in the [\ion{Ne}{2}] line follows quite faithfully
that of the radio continuum. Such resemblance between the two maps is unsurprising, as both are tracers of the thermal ionized gas. 

\begin{figure}
\resizebox{7cm}{!}{\includegraphics{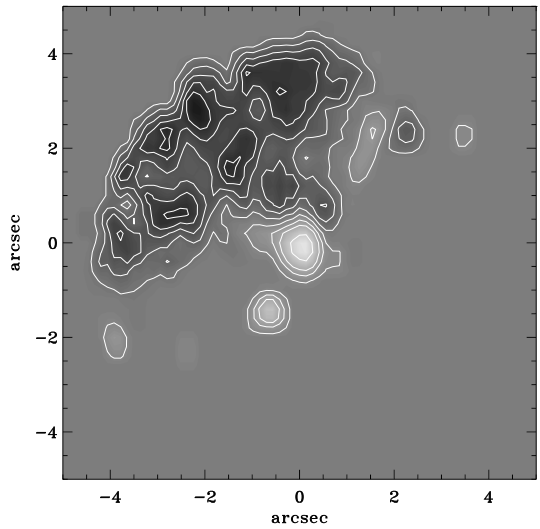}}
\caption{Spectral index distribution between 6cm and 0.7cm. Contours
levels are -0.2 to 0.7 in steps of 0.1}
\label{sp_map}
\end{figure}

To search for the existence of denser regions with different opacity through the nebula, we computed the spectral index map by comparing the 6 cm with the  0.7 cm image obtained in the first epoch. The images have comparable spatial resolution. We computed the spectral index, defined by the relation S($\nu$)~$\varpropto$~$\nu^{\alpha}$, using the maps produced with the same beam and cellsize for each pixel with a flux density of at least three times the rms noise at both the wavelengths. The resulting map, shown in Figure~\ref{sp_map}, shows evidence of the stellar wind component, characterized by 
spectral index of $\alpha$~$\approx$~0.8, and the extended halo having a much flatter radio spectrum ($\alpha$~$\approx$~-0.1), which is typical of free-free thermal emission.  However, the image shows some patchiness probably associated with local density clumps. We note the presence of a second component with spectral index of $\alpha$~$\approx$~0.9 located about 1\farcs5 to the south of the central core, in the direction 
of the southern infrared emission peak in the dust nebula. Such a component, which is clearly visible in the radio map reported in Figure 1b in \cite{Umana05}, is probably due to the presence of a dust condensation whose thermal emission contributes to the flux observed at 7 mm. \par

\section{Discussion}

\begin{figure}
\resizebox{7cm}{!}{\includegraphics{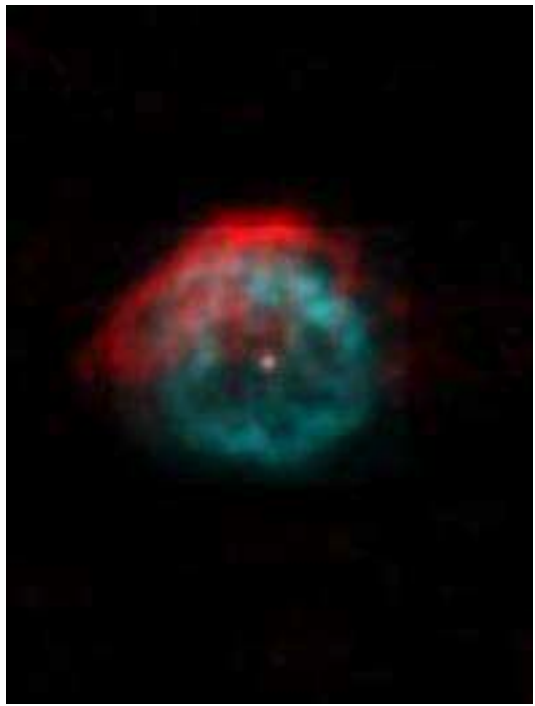}}
\caption{Composite image of IRAS~18576+0341 obtained
by superposition of radio images at 6~cm (red) and 17.65 $\mu$m (blue).
See the electronic edition of the Journal for a color version of this figure.}
\label{figcol}
\end{figure}

The most striking result of our observations is the clearly  dissimilar morphology  
of dusty and gaseous components of the nebula, which seem to have quite different spatial distribution. 
This is immediately apparent in Figure~\ref{figcol}, which shows the composite image obtained by superposition 
of the 6~cm and the 17~$\mu$m images.
The ionized nebula, in fact, extends beyond the dust shell
in the N-E direction, and the radio and [\ion{Ne}{2}] maps do not contain any indication of the symmetric shape seen in the mid-infrared continuum maps.

Different scenarios can be envisioned that could account for all 
the discussed findings and provide a complete picture of the source. 
As the continuum infrared maps seem to exclude an asymmetry in the mass loss, the asymmetry
in the distribution of the ionized gas could be ascribed to one of the following possibilities: (a)~direct photo-ionization from an external source, 
(b)~variation in the density of the interstellar medium into which the ejected material is expanding, or (c)~UV radiation from the central star leaking through a hole in the opacity of the circumstellar nebula.

The hypothesis of direct photo-ionization cannot be ruled out because of the high 
visual extinction of roughly A$_V\sim$ 28 mag \citep{ueta01} towards \object{IRAS~18576+0341}. 
The surveys available at this moment, in fact, do not have the sensitivity needed to detect
a main sequence star of spectral types O to B as it would appear at a distance of 10 kpc in this
direction. For the same reason, not even the inspection of our radio maps allow us to exclude the presence of such kind of source, although no radio emission has been detected in our map around the location of the nebula. On the basis of the analysis from Panagia $\&$ Felli (1975), in fact, the flux density expected at 6 cm from a spherically symmetric, isothermal and fully ionized stellar wind at a distance of 10 kpc is about 1.6$\times$10$^{-3}$~mJy, assuming a stellar temperature of 10$^4$~K, a terminal wind velocity of 10$^3$~km~s$^{-1}$ and a mass loss rate of 10$^{-6}~M_{\odot}~yr^{-1}$. The resulting flux is thus very close to the noise of the radio map.\par 
On the other hand, based on at the near-infrared images from the Two Micron All Sky Survey (2MASS) toward the position of \object{IRAS~18576+0341}, we found that north-east rim of the ionized cloud is facing a source designed as
\object{2MASS~19001104+0345511} in the 2MASS All-Sky Catalog of Point Sources \citep{Cutri_03} and located at position $\alpha =  19^h 00^m 11^s.05\; \delta =  +03^{\circ}45^{\prime}51^{\prime\prime}.2$. This is a distance of  4\farcs7  from the central star of IRAS~18576+0341, just in the direction where the nebula ionization appears to be strongest. However, based on the above considerations and the near-IR colours derived from the de-reddened 2MASS magnitudes, we exclude the possibility that such a source could be an early type main sequence or supergiant star able to provide the energy needed for the nebula ionization.

We can also consider the hypothesis that the ionized gas surrounding the LBV originates from the interaction of the mass outflow from the central star with a high density medium surrounding the star, such as clumps or perhaps circumstellar material. We can make an order-of-magnitude check of the plausibility of this scenario on the basis of the theory outlined by \citet{curiel_89}, who modeled the shock-ionization scenario and derived the radio continuum emission under optically thin conditions. 
In their formulation, the effective optical depth of the emitting region is directly connected to the outflow characteristics, and a simple relation exists between the radio flux density in the optically thin regime and the wind momentum rate: 

\begin{displaymath}
\frac{S(\nu)}{mJy}= 3.98\times10^{-2}\, \eta 
\left[ \frac{\nu}{5\,GHz} \right]^{0.1} 
\left[ \frac{T}{10^4\,K}\right]^{0.45} 
\left[ \frac{\dot {M}}{10^{-7}M_{\odot}yr^{-1}}\right]
\left[ \frac{v_\infty}{100\, km\,s^{-1}}\right]^{0.68}
\left[ \frac{d}{kpc}\right]^{-2}
\end{displaymath}

In this formula, $S_{\nu}$ is the radio flux observed at frequency $\nu$,
$\eta$ represents the fraction of the stellar wind that is shocked and produces the observed radio continuum emission, $\dot{M}$ is the mass loss rate, $v_{\infty}$ is the terminal velocity of the wind, $T$ is the wind temperature and $d$ the distance to the source. 
Adopting $\eta\,\approx\,$0.3  and a temperature of 10$^4$K, the density flux of about 
100 mJy observed at 6~cm can be provided by isotropic mass loss rate 
ranging from 7$\times$10$^{-3}$ to 3 $\times$10$^{-2}$ M$_{\odot}$yr$^{-1}$ for an assumed
wind velocity varying from 500 to 200 km$\,$s$^{-1}$, which are parameters values comparable 
with those derived for \object{IRAS~18676+0341} and other LBVs observed giant eruptions \citep{Smith:2005, Smith:2006a,  clark09}. Due to the many unknown parameters this estimate is only suggestive, but it shows that with typical parameters the shock emission model can produce the observed flux.

On the other hand, in absence of an external factor (source),  the ionization of the circumstellar nebula could be explained exclusively by the direct photo-ionization from the central object.
In this case, as derived in Paper~I, the Lyman continuum flux of a B0-B0.5 supergiant
can account for the measured radio flux.
This scenario, however, involves the presence of some kind of
inhomogeneity (asymmetry, anisotropies) in the circumstellar material that has not been observed in our 
mid-IR images. 
It is, in fact, difficult to understand the departures from symmetry of the radio and [\ion{Ne}{2}] images except in terms of density or optical depth variations in the cloud. \par 

The very different spatial distribution of circumstellar dust and ionized gas might reflect
an anisotropy in the same mass loss event that could have excavated the mid-IR emitting dust,
or may indicate that distinct ejection events have been collimated differently by the interaction with the circumstellar material ejected in previous phases. Such interactions could have affected the grain size distribution throughout the nebula, causing an anisotropy in the opacity of the dust grains which are competing with the gas in the absorption the UV radiation \citep{vanHoff04}.

\section{Conclusions}
We have presented new high-resolution mid-infrared and radio imaging observations of IRAS~18576+0341.
These observations provide an unprecedented view of the detailed spatial structure of the circumstellar envelope surrounding the central massive star, and highlight the different
distribution of dust and gas component in the nebula. These data allow us to perform a spatially resolved analysis of dust temperature and optical depth, and to derive the high resolution spectral index map of the observed radio emission. \par
We found that the dust properties seem to be essentially the same everywhere in the nebula and the spectral index map is consistent with optically thin free-free radiation, except for the wind component from the central star. We thus conclude that the clumpiness observed at all wavelengths has to be ascribed to local density enhancements.\par  
We also described the possible scenarios that could account for
the asymmetry in the spatial distribution of the ionized material. 
We suggest that the gas ionization could be caused by UV-radiation from an external source or
by the impinging of the mass outflow upon an obstacle like a denser circumstellar medium. Alternatively, the asymmetry could be ascribed to UV photons leaking out from the central object thought a hole in the circumstellar envelope.\par
Although we are unable to rule out any of the discussed frameworks, 
 it is worthwhile to note that the last scenario conflicts with the relative minor spatial variations of dust temperature and optical depth resulting from the analysis of our mid-IR images.
 \acknowledgments

We acknowledge partial financial support from the ASI contract
I/038/08/0 ``HI-GAL'' and from PRIN-INAF 2007.

This work is based  on observations made with the VISIR instrument on the ESO VLT telescope 
(program ID. 079.D-0748A). The Very Large Array is a facility of the
National Radio Astronomy Observatory which is operated by Associated Universities
Inc. under cooperative agreement with the National Science Foundation.

{\it Facilities:} \facility{VLT (VISIR)}, \facility{VLA}.


\begin{table}
\begin{center}
\caption{Observations Log}
\begin{tabular}{ccccccc}
\tableline\tableline
~\\
Filter     &   Date      & UT     &  Integration time   & Airmass & FWHM    \\
           &             &        &   (s)               &         & $\prime\prime$    \\
\tableline
PAH2       &  2007-07-26 & 04:56  & 700  & 1.2 & 0.27\\
PAH2$_2$   &  2007-07-26 & 05:15  & 700  & 1.3& 0.28\\
NeII       &  2007-07-19 & 06:21  & 700  & 1.4 & 0.30\\
NeII$_2$   &  2007-07-19 & 06:01  & 700  & 1.4 & 0.32\\
Q1         &  2007-07-26 & 05:48  & 180  & 1.4 & 0.38\\
\tableline
\end{tabular}
\end{center}
\end{table}

\begin{table}
\caption{Observed Mid Infrared Flux}
\label{tab_irflux}
\begin{center}
\vspace{0.3cm}
\begin{tabular}{rcr}
\hline
\hline
Filter      &F$_{\rm{star}}$  &F$_{\rm{neb}}$   \\
$\mu$m      &Jy             &Jy             \\
\hline
11.26       &0.12$\pm$0.02           &41.2$\pm$0.2           \\
11.88       &0.15 $\pm$0.02          &51.9 $\pm$0.2  \\
12.80       &0.19$\pm$0.03           &78.8 $\pm$0.1  \\
13.04       &0.17$\pm$0.03           &74.7$\pm$0.1  \\
17.65       &\dots          &224.7$\pm$0.1      \\
\hline
\end{tabular}
\label{visir-phot}
\end{center}
\end{table}

\begin{table}
\caption{Observed radio flux}
\label{tab_radio}
\begin{center}
\vspace{0.3cm}
\begin{tabular}{rrr}
\hline
\hline
Filter      &F$_{\rm{star}}$  &F$_{\rm{neb}}$   \\
 cm           &mJy             &mJy             \\
\hline
 6       & 1.3$\pm$0.04           & 103$\pm$2           \\
20       &...           & 92$\pm$2          \\
\hline
\end{tabular}
\label{visir-phot}
\end{center}
\end{table}



\begin{thebibliography}{}
\bibitem[Barlow et al.(2005)]{Barlow:2005}{Barlow}, M.,~J., Sugerman, B.~E.~K. , Fabbri, J., Meixner, M., Fisher, R.~S., Bowey, J.~E., Panagia, N., Ercolano, B., Clayton, G. C., Cohen, M., Gledhill, T.~M., Gordon,, K., Tielens, A.~G.~G.~M., \& {Zijlstra}, A.~A. 2005, \apjl, 627, L113
\bibitem[Clark et al.(2003)]{clark_03} Clark, J.~S., Egan, M.~P., Crowther, P.~A., Mizuno, D.~R., Larionov, V.~M., \& Arkharov, A. 2003, \aap, 412, 185
\bibitem[Clark et al.(2005)]{clark05} Clark, J. S., Larionov, V. M., \& Arkharov, A. 2005, \aap, 435, 239
\bibitem[Clark et al.(2009)]{clark09} Clark, J.~S.,Crowther, P.~A., Larionov, V.~M., Steele, I.~A., Ritchie, B.~W., \& Arkharov, A. A. 2009, A\&A, 507, 1555
\bibitem[Condon et al.(1999)]{Condon99} Condon, J.~J., Kaplan, D.~L., Terzian, Y. 1999, \apj S, 123, 219
\bibitem[Curiel et al.(1989)]{curiel_89}     
    Curiel, S., Rodriguez, L.~F., Bohigas, J., Roth, M., Canto, J., Torrelles, J.~M. 1989, Astro. Lett. and Comm, 27, 299
\bibitem[Cutri et al.(2003)]{Cutri_03} Cutri  R.~M.  et al. 2003, {\it The IRSA 2MASS All-Sky Catalog of Point Sources}, NASA/IPAC Infrared Science Archive, http://irsa.ipac.caltech.edu/applications/Gator
\bibitem[Fazio et al.(2004)]{fazio04} Fazio G.G. et al., 2004, ApJS, 154, 10
\bibitem[Garc\'{\i}a-Lario et al.(1997)]{garcialario} Garc\'{\i}a-Lario, P., Manchado, A., Pych, W., Pottasch, S~.R. 1997, A\&AS, 126, 479
\bibitem[Gvaramadze et al.(2010a)] {gvaramadze10a} Gvaramadze, V. V., Kniazev, A. Y., Fabrika, S., Sholukhova, O., Berdnikov, L. N., Cherepashchuk, A. M., Zharova, A. V. 2010a, MNRAS, 405, 520
\bibitem[Gvaramadze et al.(2010b)] {gvaramadze10b} Gvaramadze, V. V., Kniazev, A. Y., Fabrika, S. 2010b, MNRAS, 405, 1047
\bibitem[Humphreys \& Davidson(1994)]{HD94} Humphreys, R.~M., Davidson, K. 1994, PASP, 106, 1025
\bibitem[Hrivnak et al.(2000)]{Hrivnak00} Hrivnak, B.~J., Volk, K., Kwok, S. 2000, \apj, 535, 275.
\bibitem[Jim\'enez et al.(2010)]{Jimenez10}Jim\'enez-Esteban, F.~M., Rizzo, J.~R. \& Palau, A. 2010, \apj, 713,429
\bibitem[Kotak \& Vink(2006)]{Kotak:2006} Kotak, R. \& Vink,  J.~S. 2006, A\&A, 460, L5
\bibitem[Lagage et al.(2004)]{Lagage2004} Lagage, P.~O., Pel et al. 2004, The Messenger, 117, 12.
\bibitem[Lang et al.(2005)]{Lang05} Lang, C.~C., Johnson, K~ E., Goss, W.~M., Rodr\'{\i}guez, L.~F. 2005, AJ, 130, 2185
\bibitem[Lang et al.(1999)]{Lang99} Lang, C.~C., Figer, D.~F., Goss, W.~M., \& Morris, M. 1999, AJ, 118, 2327 
\bibitem[Nota et al.(1997)]{not97} Nota A., Smith, L., Pasquali, A., Clampin, M., \&  Stroud, M. 1997 \apj, 486, 338
\bibitem[Pasquali \& Comeron(2002)]{Pasquali_02} Pasquali, A. \& Comeron, F. 2002, A\&A, 382, 1005
\bibitem[Rieke et al.(2004)]{rieke04} Rieke G.H. et al., 2004, ApJS, 154, 25
\bibitem[Smith(2005)]{Smith:2005} Smith, N. 2005, MNRAS, 357, 1330
\bibitem[Smith \& Hartigan(2006)]{Smith:2006a} Smith, N. \& Hartigan,~P. 2006, \apj, 638, 1045
\bibitem[Smith \& Owocki(2006)]{Smith:2006} Smith, N. \& Owocki, S.~P. 2006, \apjl, 645, L45
\bibitem[Sloan et al.(2003)]{Sloan03}Sloan, G.~C., Kraemer, K.~E., Price, S.~D., Shipman, R.~F. 2003, \apss, 147, 379
\bibitem[Ueta et al.(2001)]{ueta01}   Ueta, T., Meixner, M., Dayal, A., Deutsch, L.~K., Fazio, G.~G., Hora, J.~L., Hoffmann, W.~F. 2001, \apj, 548, 1028
\bibitem[Umana et al.(2005)]{Umana05} Umana, G., Buemi, C.~S., Trigilio, C., Leto, P. 2005, A\&A, 437, L1
\bibitem[Umana et al.(2009)]{Umana09}   Umana, G., Buemi, C.~S., Trigilio, C., Hora, J.~L., Fazio, G.~G., Leto, P. 2009, \apj, 694, 697
\bibitem[Umana et al.(2010)]{Umana10}  Umana, G., Buemi, C.~S., Trigilio, C.,\& Leto, P. 2010, \apj, 718, 1036
\bibitem[van Hoff et al.(2004)]{vanHoff04} van Hoff, P., A., M., Weingartner, J., C., Martin, P., J., Volk, K., and Ferland, G., J., 2004, \mnras, 350, 1330
\bibitem[Wachter et al.(2010)]{wachter10}Wachter, S., Mauerhan, Jon C., Van Dyk, Schuyler D., Hoard, D. W., Kafka, S., Morris, P. W. 2010, AJ, 139, 2330
\bibitem[White(2000)]{white00} White, S.~M. 2000, \apj, 539, 851

\end{thebibliography}
\end{document}